\documentclass[aps,prd,secnumarabic,nobibnotes,twocolumn,superscriptaddress]{revtex4-1}
\usepackage{amsfonts}
\usepackage{mathrsfs}
\usepackage{amsmath}
\usepackage{color}
\usepackage{natbib}
\usepackage{graphicx}
\usepackage{bm}
\usepackage{amssymb}
\usepackage{xspace}
\usepackage{epstopdf}
\usepackage{dcolumn}
\usepackage{multirow}
\usepackage[colorlinks=true, letterpaper=true, pdfstartview=FitV, linkcolor=blue, citecolor=blue, urlcolor=blue]{hyperref}
\usepackage{wrapfig}

\makeatletter

\newcommand{\Rmnum}[1]{\expandafter\@slowromancap\romannumeral #1@}
\makeatother

\begin{document}
\title{Fully spin-polarized double-Weyl fermions with type-III dispersion in quasi-one dimensional materials X$_2$RhF$_6$ (X=K, Rb, Cs)}

\author{Lei Jin}
\affiliation{School of Materials Science and Engineering, Hebei University of Technology, Tianjin 300130, China.}

\author{Xiaoming Zhang}
\email{zhangxiaoming87@hebut.edu.cn}
\affiliation{School of Materials Science and Engineering, Hebei University of Technology, Tianjin 300130, China.}

\author{Ying Liu}
\affiliation{School of Materials Science and Engineering, Hebei University of Technology, Tianjin 300130, China.}

\author{Xuefang Dai}
\affiliation{School of Materials Science and Engineering, Hebei University of Technology, Tianjin 300130, China.}

\author{Liying Wang}
\affiliation{Tianjin Key Laboratory of Low Dimensional Materials Physics and Preparation Technology, School of science, Tianjin University, Tianjin 300354, People's Republic of China.}

\author{Guodong Liu}
\email{gdliu1978@126.com}
\affiliation{School of Materials Science and Engineering, Hebei University of Technology, Tianjin 300130, China.}

\begin{abstract}
Double-Weyl fermions, as novel topological states of matter, have been mostly discussed in nonmagnetic materials. Here, based on density-functional theory and symmetry analysis, we propose the realization of fully spin-polarized double-Weyl fermions in a family ferromagnetic materials X$_2$RhF$_6$ (X= K, Rb, Cs). These materials have the half-metal ground states, where only the bands from the spin-down channel present near the Fermi energy. The spin-down bands form a pair of triply degenerate nodal points (TDNPs) if spin-orbit coupling (SOC) is not included. Under SOC, one TDNP splits into two double-Weyl points featuring quadratic dispersion along two momentum direction, and they are protected by the three-fold rotation ($C$$_3$) symmetry. Unlike most double-Weyl semimetals, the Weyl points proposed here have the type-III dispersion with one of the crossing bands being saddle-shaped. An effective model is constructed, which describes well the nature of the Weyl points. These Weyl points are fully spin-polarized, and are characterized with double Fermi arcs on the surface spectrum. Breaking $C$$_3$ symmetry by lattice strain could shift one double-Weyl point into a pair of type-II single-Weyl points. The X$_2$RhF$_6$ materials proposed here are excellent candidates to investigate the novel properties of type-III double-Weyl fermions in ferromagnetic system, as well as generate potential applications in spintronics.

\end{abstract}
\maketitle

\section{Introduction}
The discoveries of graphene and three-dimensional (3D) topological insulators have sparked the research enthusiasm on topological aspects of matter~\cite{1,2,3,4,5}. In recent years, Weyl semimetals (WSMs) are believed to open new era in the field and have received worldwide attentions~\cite{6,7,8,9,10}, partially because they provide a wonderful feasibility of realizing the phenomenon of high-energy physics in the condensed matter scale. In a WSM, the low-energy excitations, referred as Weyl fermions, can be viewed as separated "magnetic monopole" (the sources/sinks of the Berry curvature), which only exists in the traditional model but has not been identified in nature as a fundamental particle. Unlike topological insulators, the bulk band structure for a WSM is metallic; however, the nontrivial band topology is also applied. Associated with the nontrivial band topology, WSMs are characterized with novel Fermi arc electron states on the surface~\cite{11,12,13,14,15}. The Weyl fermions in the bulk and the Fermi arc states on the surface are believed to bring a wide range of exotic properties such as ultrahigh mobility, quantum Hall effect, chiral magnetic effect and negative longitudinal magnetoresistance~\cite{16,17,18,19,20}.

\begin{figure}
\includegraphics[width=8.8cm]{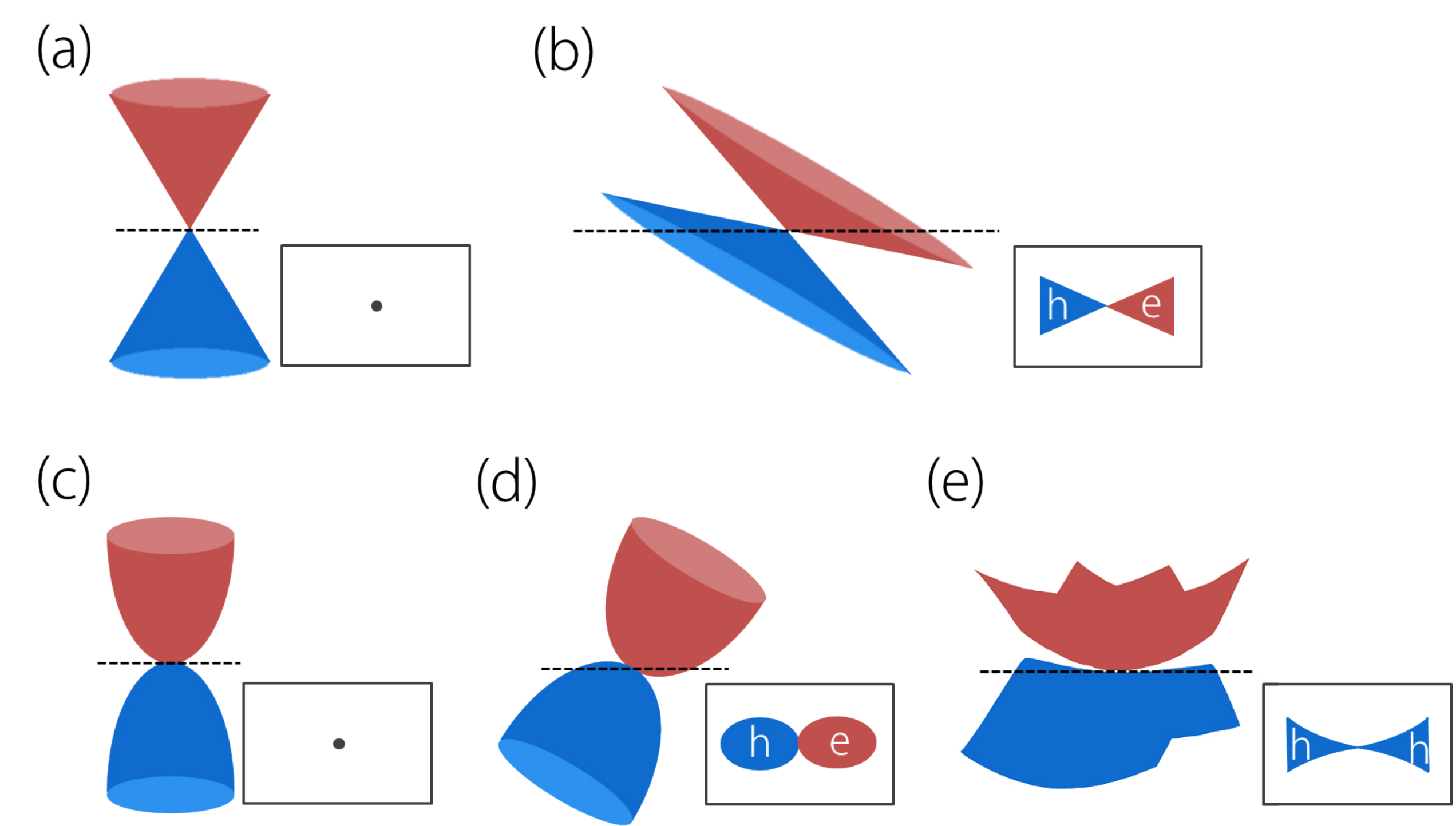}
\caption{Schematic for (a) type-I, (b) type-II Weyl semimetals characterized by linear dispersion. The insets of (a) and (b) show the type-I and type-II Weyl semimetals with a point-like Fermi surface and the Fermi surface of coexisting hole-like and electron-like states, respectively. Schematic for (c) type-I, (d) type-II and (e) type-III Weyl semimetals characterized by quadratic dispersion. (c) and (d) show the type-I and type-II Weyl semimetals with a point-like Fermi surface and the Fermi surface of coexisting hole-like and electron-like states, respectively. (e) type-III Weyl semimetal has a unique Fermi surface connecting of two hole-like pockets (or two electron-like pockets) from the opposite quadratic tilting term).
\label{fig1}}
\end{figure}

WSMs can be classified into distinct categories based on different dispersions of the crossing bands. Most commonly, WSMs are characterized by linear dispersion of the Weyl points as shown by the cases in Fig.~\ref{fig1}(a) and (b). These Weyl points have the chiral charge of $\pm$1~\cite{6,7,16,17}. Besides, when additional rotation symmetry exists ($C$$_3$, $C$$_4$, or $C$$_6$) in the system, more exotic Weyl states can be formed. In such a case, the Weyl points can show quadratic (or cubic) dispersion in two momentum directions, which lead to the formation of higher charge monopoles for the Berry connection~\cite{18,19,20,21,22}. For example, quadratic crossing between two bands constitute the so-called double Weyl points, which have the chiral charge of $\pm$2. Figure.~\ref{fig1}(c) - (e) shows typical band dispersions for double Weyl states. In addition, different slopes of crossing bands differentiate WSMs into the categories of type-I and type-II. Type-I WSMs are characterized with a point-like Fermi surface~\cite{17,23,24,25}, because of the well separated hole-like and electron-like states [see Fig.~\ref{fig1}(a)]. In type-II WSMs ~\cite{26,27,28}, the Fermi surface shows the geometry of coexisting hole-like and electron-like states, because of the completely titled Weyl cone [see Fig.~\ref{fig1}(b)]. Type-II WSMs are proposed to exhibit distinct properties with the type-I ones~\cite{29,30,31,32,33,34}. Similarly, double Weyl states can also show type-I and type-II cases, as shown in Fig.~\ref{fig1}(c) and (d), respectively. In addition to these types, Li \emph{et al}. proposed the type-III category of WSMs quite recently~\cite{35}. As shown in Fig.~\ref{fig1}(e), the formation of a type-III WSM requires one of the crossing bands has the saddle-shape, which could induce a unique Fermi surface connecting of two electron-like or (hole-like) pockets.

To form a WSM, at least one of the inversion symmetry (IS) and the time reversal symmetry (TRS) needs to be broken. IS-breaking WSMs have been proposed in many materials and some have been realized in experiments~\cite{23,24,25,26,27,28,29,30,31,32,33,36,37}. However, the candidates for TRS-breaking WSMs are more scarce, where a few magnetically-ordered materials including Y$_2$Ir$_2$O$_7$~\cite{6}, HgCr$_2$Se$_4$~\cite{38}, RAlGe (R = rare earth) compounds~\cite{39,40}, and some Heuslers are such examples~\cite{41,42}. Especially, it is reported in Co$_3$Sn$_2$S$_2$ that magnetic Weyl phase can enable unusually large anomalous Hall conductivity~\cite{43,44}, which was not observed in IS-breaking WSMs. Comparing with traditional WSMs, the nonlinear energy dispersions in double-Weyl semimetals (D-WSMs) are potential to induce several unique characteristics, such as striking non-Fermi-liquid behaviors and anisotropic thermoelectric properties~\cite{45,46,47,48,49}. To date, only a few D-WSMs are reported, including HgCr$_2$Se$_4$~\cite{38}, SrSi$_2$~\cite{50}, Eu$_5$Bi$_3$~\cite{51}, FeSi ~\cite{52}, CoSi ~\cite{53}, Blue Phosphorene Oxide~\cite{54}, and (TaSe$_4$)$_2$I~\cite{35}. In these examples, only HgCr$_2$Se$_4$ and Eu$_5$Bi$_3$ are reported to show magnetic ordering~\cite{38,51}. Unfortunately, experimental observations of the magnetic double-Weyl states in them are still in difficulty currently. Thus, it remains significant to explore TRS-breaking D-WSMs with magnetic ordering.

	In current work, we report a novel family of TRS-breaking D-WSMs in quasi-one dimensional compounds X$_2$RhF$_6$ (X=K, Rb, Cs). These materials naturally show the ferromagnetic ordering and have half-metallic band structure. Taking K$_2$RhF$_6$ as a concrete example, we find the band structure in the spin-down channel shows a pair of triply degenerate nodal points (TDNPs) along the $\Gamma$-A path when spin-orbit coupling (SOC) is not considered. When SOC is include, the TDNPs split into pairs of double-Weyl points near the Fermi level. Remarkably, the double-Weyl points show the type-III dispersion, because one of the crossing bands is saddle-shaped. Such type-III D-WSMs have not been proposed in magnetic system previously. Clear Fermi arc surface states for the double-Weyl points are identified in the material. The effective model for the double-Weyl points is constructed. In addition, we show a double-Weyl point would transform into a pair single-Weyl points with type-II dispersion under uniaxial lattice strain breaking the $C$$_3$ rotation symmetry. Our results show that X$_2$RhF$_6$ compounds can play good material platforms to investigate the fundamental physics of type-III double-Weyl fermions in fully spin-polarized ferromagnets.

\section{Computational methods}

The first principles calculations were performed by using the Vienna ab initio Simulation Package (VASP), based on density-functional theory (DFT)~\cite{55,56}. For the exchange correlation-potential, we adopt the generalized gradient approximation (GGA) of the Perdew-Burke-Ernzerhof (PBE) functional~\cite{57}. The interaction between electrons and ions was modeled by using the projector augmented wave (PAW) method~\cite{58}.The cutoff energy was set as 500 eV. A $\Gamma$-centered $k$-mesh of 11$\times$11$\times$13 was used for the Brillouin zone (BZ) sampling. During lattice optimization, energy and force convergence criteria were set as $10^{-6}$ eV and 0.01 eV $\AA^{-1}$, respectively. The topological features of surface states were calculated based on the maximally localized Wannier functions~\cite{59,60}, realized by using the WANNIERTOOLS package~\cite{61}.

\section{Crystal structure and magnetic property}

\begin{figure}
\includegraphics[width=8.8cm]{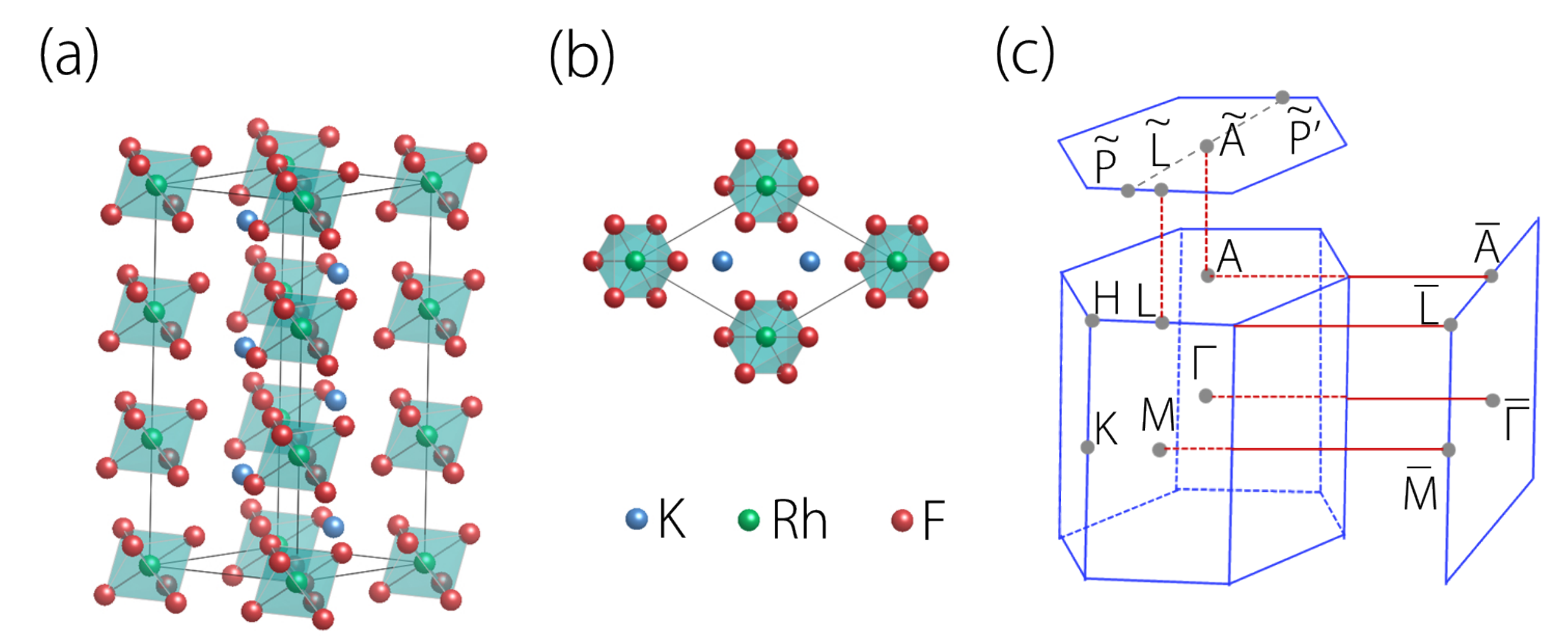}
\caption{(a) 1$\times$1$\times$3 supercell of K$_2$RhF$_6$. One Rh atom and adjacent six F atoms form an octahedron. Crystal of K$_2$RhF$_6$ contains octahedron RhF$_6$ chains along the $c$ axis. The K atoms occupy the middle regions among the chains. (b) Top view of crystal structure of K$_2$RhF$_6$. (c) The bulk, (001) and (010) surfaces Brillouin zone of K$_2$RhF$_6$.
\label{fig2}}
\end{figure}

The series of X$_2$RhF$_6$ (X=K, Rb, Cs) compounds are all existing materials, and their crystal structures and physical properties have already been well characterized since the 1950s~\cite{62}. These compounds share the same hexagonal lattice with the space group \emph{P$\bar{3}$m1} (No. 164). Taking K$_2$RhF$_6$ compound as an example, the crystal structure is shown in Fig.~\ref{fig2}(a) and (b). One Rh atom bonds with six F atoms, and constitute the RhF$_6$ local structure. Such local structure stacks along the $c$-axis and forms RhF$_6$ chains, while K atoms occupy the middle regions among the chains. This clearly manifests K$_2$RhF$_6$ is a quasi-one dimensional material. During our computations, the optimized lattice constants of K$_2$RhF$_6$ compound yield to be a = b = 5.85 \AA and c = 4.76 \AA, matching well with the experimental ones (a = b = 5.74 \AA and c = 4.65 \AA)~\cite{62}. In the optimized structure, K, Rh and F atoms occupy the 2$d$ (0.3333, 0.6667, 0.7), 1$a$ (0, 0, 0) and 6$i$ (0.17, -0.17, 0.22) Wyckoff sites, respectively. We use the optimized lattice structure in the following calculations. In addition, the conclusions in this work will not change if the experimental structure is employed.

Former experimental investigations have observed the appearance of magnetic ordering in K$_2$RhF$_6$ compound~\cite{62}. Our calculations also give out consistent results. We compare the total energies between the spinless and spin-polarized states in K$_2$RhF$_6$ compound, and find that the spin-polarized one is lower in 88.37 meV per unit cell. In K$_2$RhF$_6$ compound, we find the magnetic moments are mostly contributed by the Rh atom. Further, we have considered typical magnetization directions in the hexagonal lattice, including [1 0 0], [0 1 0], and [0 0 1]. Our calculations show that the magnetization along the [0 0 1] direction possesses the lowest energy. Besides the ferromagnetic ordering, we have also considered several antiferromagnetic configurations in different sizes of supercells~\cite{63}. Our results show the antiferromagnetic states are about 16 meV higher than the ferromagnetic state. Therefore, it has been confirmed that, K$_2$RhF$_6$ compound has the ferromagnetic ground state, and the easy magnetization direction is along the $c$-axis. It is very interesting to note that, K$_2$RhF$_6$ compound has an integer magnetic moment of 1.0 $\mu$$_B$ per unit cell.

\section{Topological band structure}

\begin{figure}
\includegraphics[width=8.8cm]{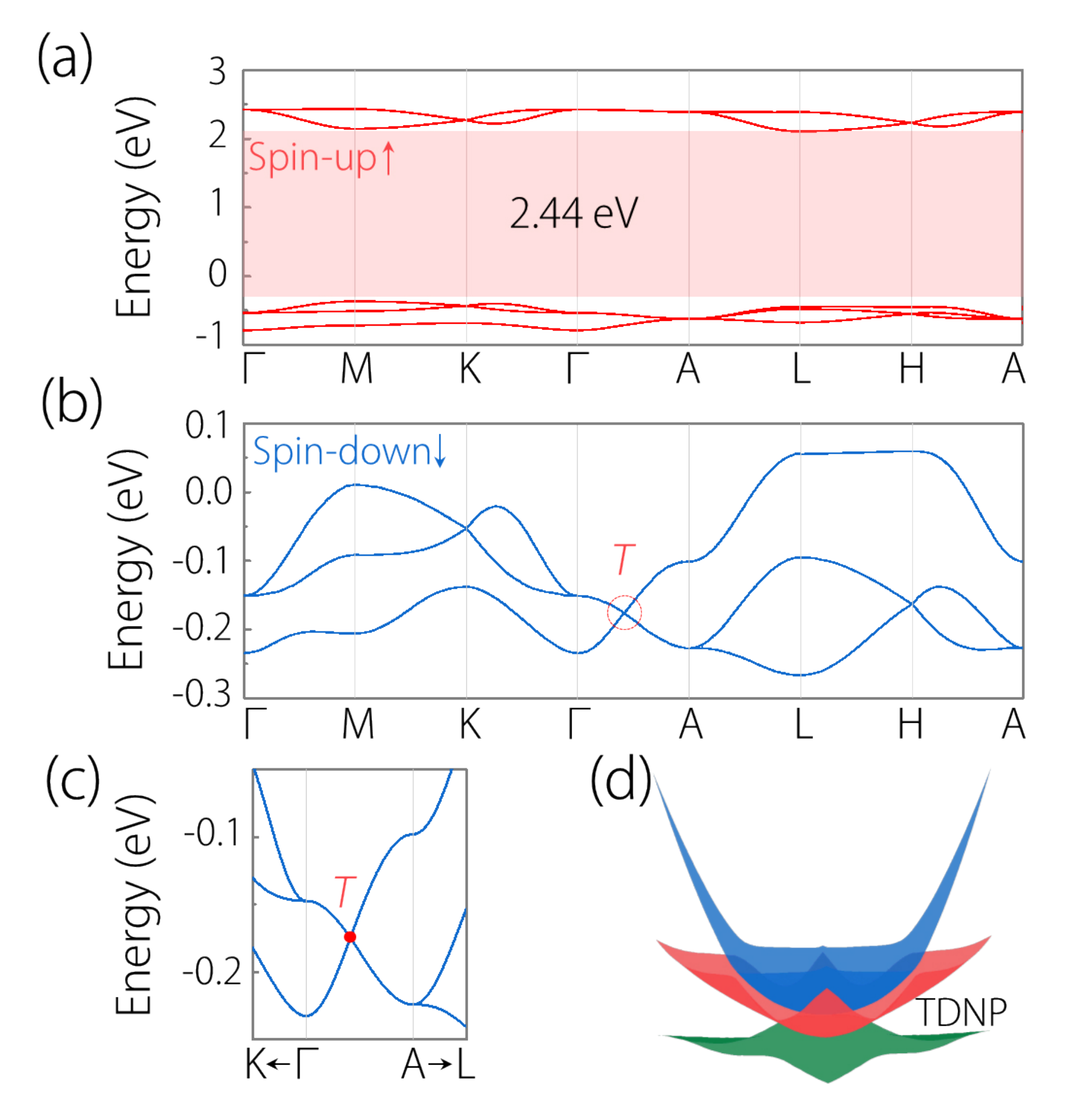}
\caption{The electronic band structures of K$_2$RhF$_6$ compound in the absence of SOC. (a) Spin-up band structure, exhibiting an insulating character with a big band gap of 2.44 eV. (b) Enlarged spin-down band structure, showing a metallic character with bands crossing the Fermi level. $T$ indicates triply degenerate nodal points. (c) Enlarged view of the band structure around the $T$ point. (d) The three-dimensional plot of band dispersions near the $T$ point.
\label{fig3}}
\end{figure}

We first discuss the electronic band structure of K$_2$RhF$_6$ compound without considering SOC. In Fig.~\ref{fig3}(a) and (b), we show the spin-resolved band structures of K$_2$RhF$_6$ compound. The band structures show two crucial features. For the first one, we find band structures in the spin-up and spin-down channels show distinct conducting behaviors. As shown in Fig.~\ref{fig3}(a), the spin-up bands show a large band gap of 2.44 eV around the Fermi level, manifesting the insulating character. However, as shown in Fig.~\ref{fig3}(b), the spin-down bands exhibit a definitely metallic character with several bands crossing the Fermi level. These results suggest K$_2$RhF$_6$ compound is an excellent half-metal phase, where the fully spin-polarized conduction electrons can be realized. By examining the density of states~\cite{63}, we find the states near the Fermi level are mostly contributed by the $d$$_{z^2}$, $d$$_{x^2-y^2}$ and $d$$_{xy}$ orbitals of Rh atom. For the second feature, we find there shows a band crossing point in $\Gamma$-A path in the spin-down channel [see Fig.~\ref{fig3}(b)].We show the enlarged band structure near the crossing point in Fig.~\ref{fig3}(c). One can clearly observe that it is formed by the crossing among three bands (one doubly-degenerated band and one singlet). These results indicate this crossing in fact forms a TDNP. In Fig.~\ref{fig3}(d), we display the 3D plotting of band structure for these bands. Similar TDNPs have also been observed in many nonmagnetic materials with hexagonal lattice~\cite{64,65,66,67,68}.

\begin{figure}
\includegraphics[width=8.8cm]{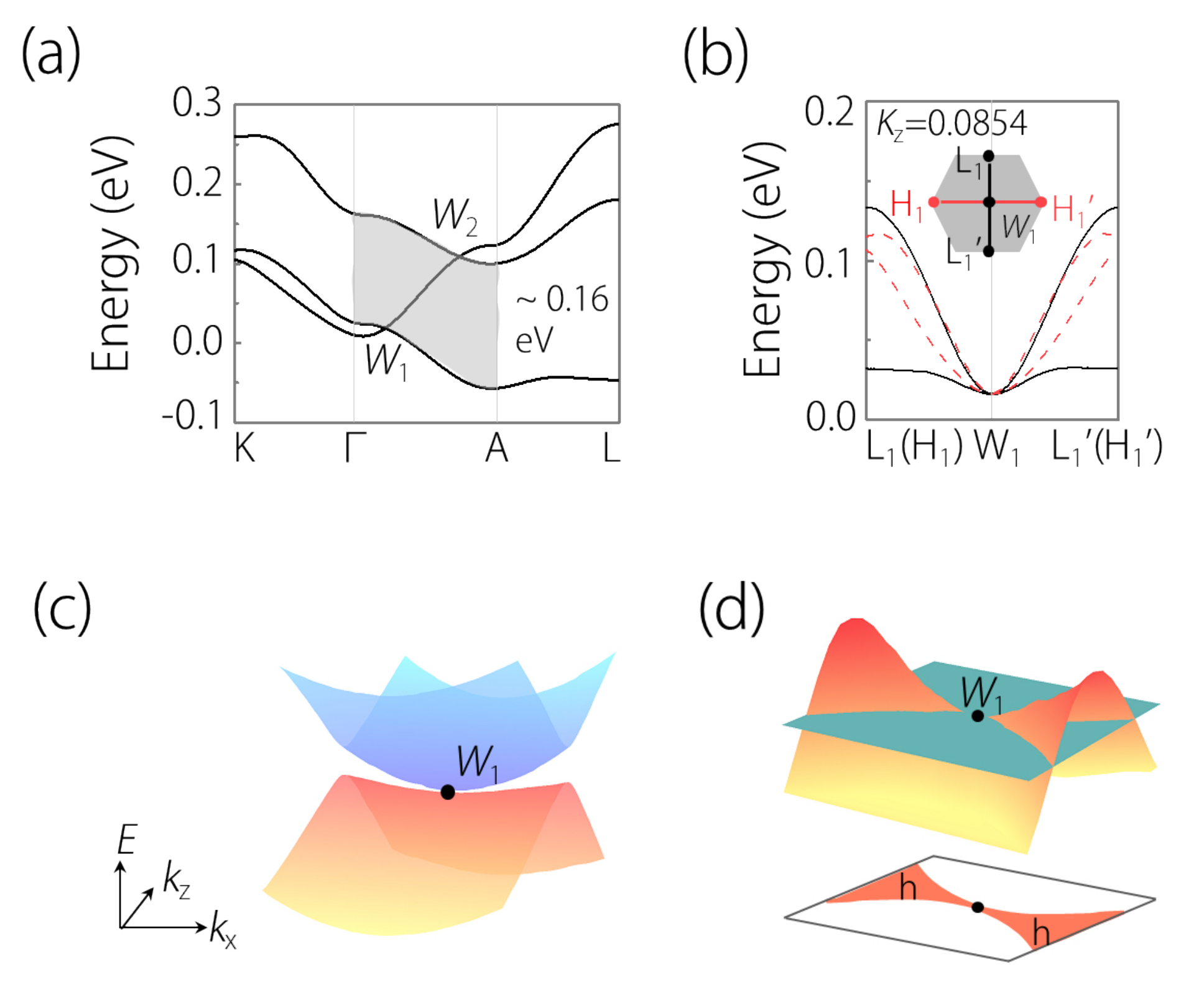}
\caption{(a) The electronic band structure of K$_2$RhF$_6$ with SOC under out-of-plane [001] magnetization. $W$$_1$ and $W$$_2$ indicate double Weyl points. (b) In horizontal plane ($k$$_z$=0.0854), quadratic energy dispersions of $W$$_1$ Weyl cone under SOC. (c) The three-dimensional plots of the energy dispersions near $W$$_1$ in the $k$$_x$-$k$$_z$ plane. (d) Fermi surface of type-III double-Weyl point $W$$_1$, which has two contacted hole pockets. The green surface indicates the equal energy surface at $W$$_1$.
\label{fig4}}
\end{figure}

Noticing K$_2$RhF$_6$ compound has considerable atomic weight, the intrinsic SOC effects would pronouncedly influence the band structures. This has been confirmed by our DFT calculations. As shown in Fig.~\ref{fig4}(a), under SOC we find the band structures near the Fermi level have experienced great changes. Especially, the doubly-degenerated band for the TDNP splits away from each other with the gap of about 0.16 eV. This band splitting gives rise two new band crossing points in the $\Gamma$-A path, as labeled as $W$$_1$ and $W$$_2$ in Fig.~\ref{fig4}(a). These band crossing points both have the double degeneracy, known as Weyl points. We notice both Weyl points locate very near the Fermi level, where $W$$_1$ is at 0.02 eV and $W$$_2$ at 0.1 eV above the Fermi level. Since $W$$_1$ locates much closer to the Fermi level, in the following we pay special attention on $W$$_1$. After a careful examination on the band structures nearby, it is interesting to find that $W$$_1$ manifest quadratic dispersion relations along two momentum directions (H-H' and L-L'), as shown in Fig.~\ref{fig4}(b). This suggests $W$$_1$ is a double-Weyl point.

In Fig.~\ref{fig4}(c), we show the 3D plotting of the band dispersions around $W$$_1$. It is interesting to note that, one of the crossing bands possesses a saddle-like dispersion. The constant energy surface at $W$$_1$ is shown in Fig.~\ref{fig4}(d), which clearly shows the connecting between two hole-like pockets. Such band dispersion is drastically different with most D-WSMs proposed previously~\cite{38,50,51,52,53,54}, where none saddle-like dispersion was found. Such novel double-Weyl points were termed as type-III recently~\cite{35}, with distinct properties from type-I and type-II ones proposed. The presence of double-Weyl points in K$_2$RhF$_6$ compound are protected by the three-fold rotation ($C$$_3$) symmetry in the $\Gamma$-A path.

Further insights on the double-Weyl points in K$_2$RhF$_6$ can be obtained from the effective \emph{k$\cdot$p} model on the basis of symmetry analysis. Corresponding to the $\Gamma$-A path, the little group is $C$$_3$. The two crossing bands for $W$$_1$ have different 1D irreducible representations $\Gamma$$_4$ and $\Gamma$$_6$, respectively. In the absence of SOC, due to the decoupling between spin-up and spin-down channels, each channel can be regarded as a spinless case. Therefore, the time reversal symmetry for each one can be given by $\mathcal{T}=\mathcal{K}$, with $\mathcal{K}$ is the complex conjugation operator. Furthermore, the little group of path $\Gamma$-A belongs to $C_{3v}$ which has two generators: a three-fold rotation $C_{3z}$ and a mirror $M_y$. The two degenerate bands corresponds to the $E$ representation, while the single band corresponds to $A_1$. Expanding up to the quadratic $k$ order, the effective Hamiltonian is given by

\begin{eqnarray}
  \mathcal{H} &=& \mathcal{H}_{1}+\mathcal{H}_{2},
\end{eqnarray}

\begin{eqnarray}
  \mathcal{H}_{1} &=& \left[\begin{array}{ccc}
                          M_0 & ivk_x & -ivk_y \\
                          -ivk_x & M+\gamma k_xk_z & 0 \\
                          ivk_y & 0 & M-\gamma k_xk_z
                        \end{array}\right],
\end{eqnarray}

\begin{equation} \label{eqn2}
   \begin{split}
   \mathcal{H}_{2} = \lambda_{1}[\alpha(k_x^2-k_y^2)+\beta k_x k_z] + \lambda_{4}(2\alpha k_x k_y-\beta k_y k_z)  \\
   + \lambda_{6}(\gamma k_yk_z+\eta k_x k_y) + \lambda_{0}(Ak_x^2+Bk_y^2+Ck_z^2),
  \end{split}
\end{equation}
where $M_0,\ M,\ \alpha,\ \beta,\ \gamma,\ \eta,\ v$ are real parameters, $\lambda_{i}$($i$ = 1, 4, 6) are the three symmetric Gell-Mann matrices, $\lambda_{0}$ stand for a matrix:

{\begin{eqnarray}
\lambda_{1}=\begin{bmatrix}
0 & 1 & 0 \\
1 & 0 & 0 \\
0 & 0 & 0
\end{bmatrix},  & \lambda_{4} = \begin{bmatrix}
0 & 0 & 1 \\
0 & 0 & 0 \\
1 & 0 & 0
\end{bmatrix},  & \lambda_{6} = \begin{bmatrix}
0 & 0 & 0 \\
0 & 0 & 1 \\
0 & 1 & 0
\end{bmatrix},
\end{eqnarray}

{\begin{eqnarray}
 \lambda_{0} = \begin{bmatrix}
1 & 0 & 0 \\
0 & -1 & 0 \\
0 & 0 & -1
\end{bmatrix}.
\end{eqnarray}

In the presence of SOC, mirror operation $M_y$ is broken, only $C_{3z}$ is remained. There appears two Weyl points along $\Gamma$-A. We expend the model at the crossing bands, then deriving the corresponding effective model as

\begin{eqnarray}
  \mathcal{H}_{W_1} &=& (M_0+Ak_\|^2+v_0k_z)\sigma_0+(\gamma k_\|^2+vk_z)\sigma_z
\end{eqnarray}
where $k_\|=\sqrt{k_x^2+k_y^2}$. We can clearly find the double-Weyl signature in the Hamiltonian. In addition, when $|A|>|\gamma|$, it shows a type-III Weyl node.

\begin{figure}
\includegraphics[width=8.8cm]{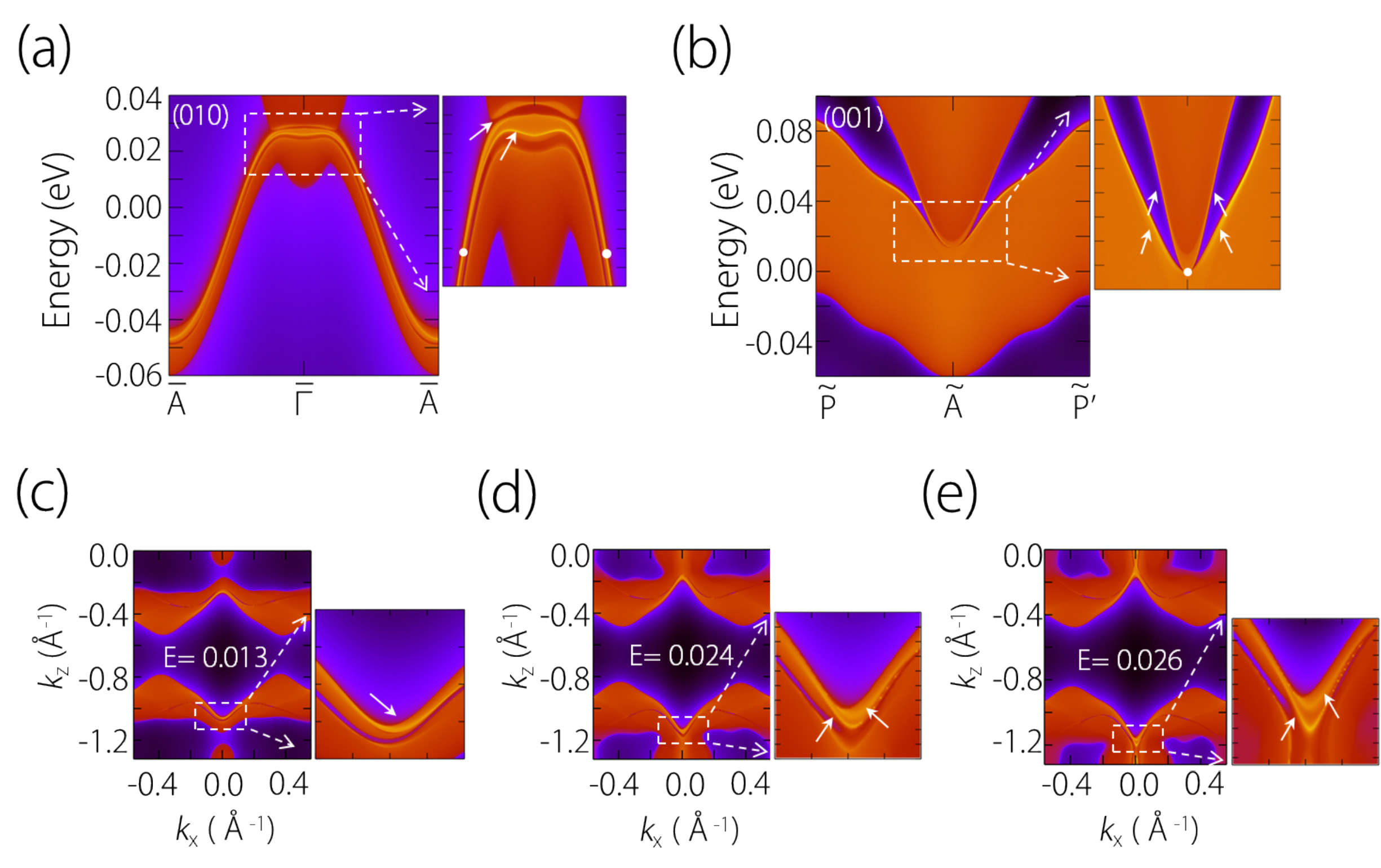}
\caption{(a) Projected spectrum of the (010) surfaces of K$_2$RhF$_6$. In enlarged plot of the surface state, there are two pieces of Fermi arcs (pointed by the white arrows) connecting the pair of $W$$_1$ (the white circles). (b) Projected spectrum of the (001) surfaces of K$_2$RhF$_6$. In enlarged plot of the surface state, two pieces of Fermi arcs (pointed by the white arrows) on both sides of the Weyl point $W$$_1$ (the white circle). The constant energy slices corresponding to (010) surface at (c) $E$= 0.013 eV, (d) $E$= 0.024 and (e) $E$= 0.026 eV. The arrows point the Fermi arcs. As the energy increases, the two Fermi arcs become more clear.
\label{fig5}}
\end{figure}

A key signature of a Weyl point is the presence of Fermi arc surface states. In traditional WSMs, Weyl points are characterized with singlet of Fermi arc surface states. However, double-Weyl points would show two pieces of Fermi arcs connecting projected images of the two bulk nodes of opposite charges~\cite{38,50,51,52,53,54}. In Fig.~\ref{fig5}(a), we show the projected band spectrum on the (010) surface of K$_2$RhF$_6$ compound. As shown in the enlarged region in Fig.~\ref{fig5}(a), it clearly reveals two pieces of Fermi arcs (pointed by the white arrows) connecting the pair of $W$$_1$ (the white circles). In addition, we have also calculated the Fermi surface plots near $W$$_1$ (at $E$ = 0.013 eV, $E$ = 0.024 eV and $E$ = 0.026 eV), as shown in Fig.~\ref{fig5}(c), (d) and (e). In these energy levels, the profile for the Fermi arcs can also be identified. Besides the (010) surface projection, we also calculate the band spectrum on the (001) surface, and the results are shown in Fig.~\ref{fig5}(b). In such surface spectrum, one can still observe two pieces of Fermi arcs on both sides of the Weyl points. The clear Fermi arc states on different surfaces can greatly facilitate their detections in future experiments.

\begin{figure}
\includegraphics[width=8.8cm]{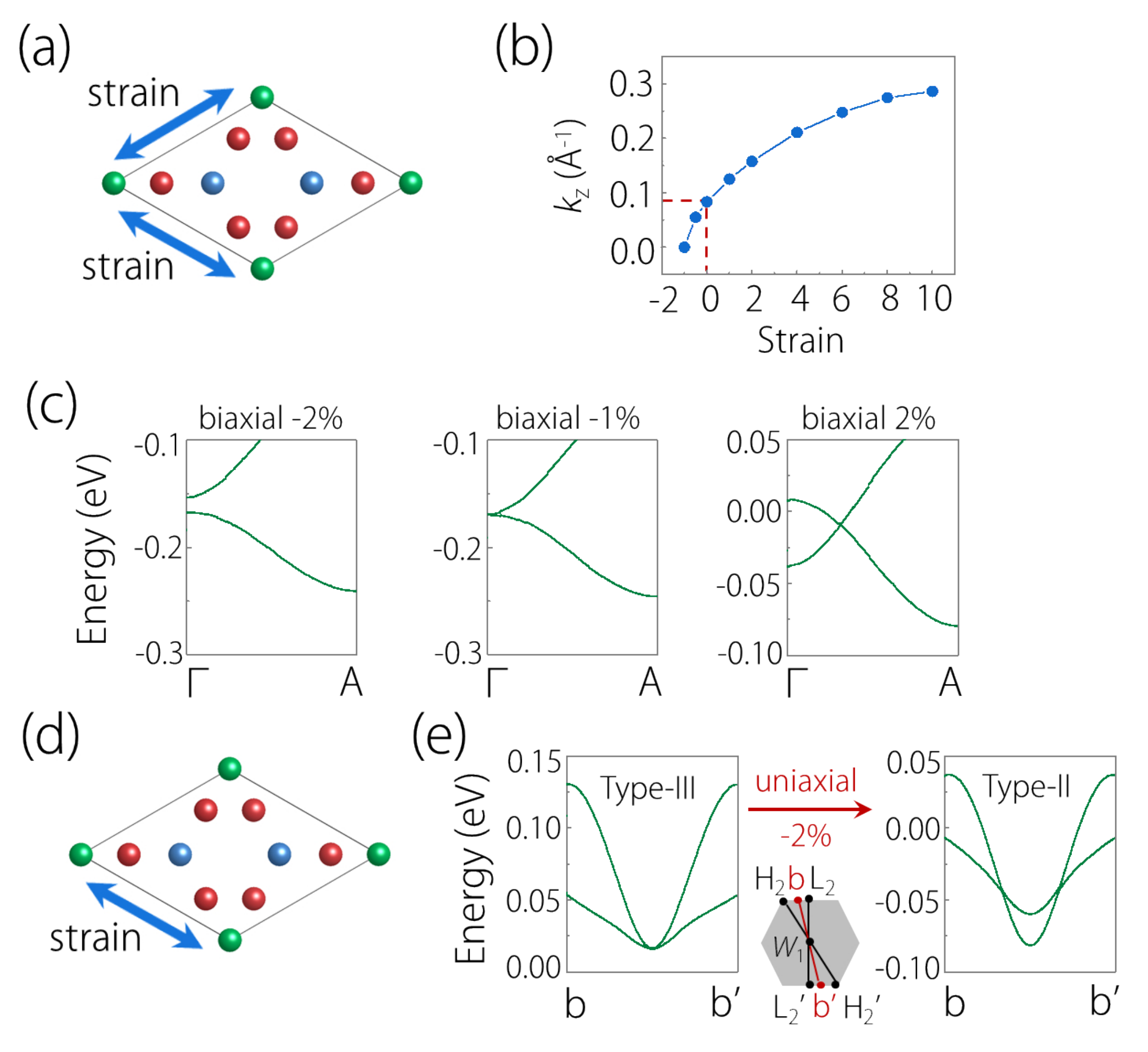}
\caption{(a) Schematic figure of K$_2$RhF$_6$ under biaxial in-plane strain. (b) Change of $W$$_1$ position under biaxial strain. The vertical coordinates show the positions of the double-Weyl point $W$$_1$ on the $\Gamma$-A path with respect to (0, 0, $k$$_z$). When compressive strain is larger than 1$\%$, the $W$$_1$ annihilates. (c) Electronic band structures of K$_2$RhF$_6$ along the $\Gamma$-A path under -2$\%$, -1$\%$ and 2$\%$ biaxial strain. (d) Schematic figure of K$_2$RhF$_6$ under uniaxial strain along \emph{a/b} direction. (e) Under -2$\%$ uniaxial strain, topological phase transitions from type-III double-Weyl point to type-II single-Weyl points. The b-b' path locates on $k$$_z$=0.07 plane.
\label{fig6}}
\end{figure}

As pointed above, the double-Weyl points in K$_2$RhF$_6$ are protected the $C$$_3$ rotation symmetry in the $\Gamma$-A path. They can be robust against weak perturbations if $C$$_3$ is preserved. Nevertheless, a stronger perturbation may pull the crossing bands apart, during which the double-Weyl points annihilate. Such a condition can be realized by strain engineering. Here, we apply a biaxial strain in the \emph{a-b} plane [see Fig.~\ref{fig6}(a)], under which the $C$$_3$ rotation symmetry preserves. We find the double-Weyl points can retain under tensile strains less than 10$\%$. With increasing the tensile strain, the double-Weyl points $W$$_1$ move away from the $\Gamma$ point in the $\Gamma$-A path. Figure.~\ref{fig6}(b) shows the distance between $W$$_1$ and the $\Gamma$ point in momentum space under biaxial strain. However, under compressive strains larger than 1$\%$, the double-Weyl points would annihilate because the crossing bands are separated. The band structures near $W$$_1$ at typical strains are shown in Fig.~\ref{fig6}(c).

In addition, we find one double-Weyl point can further transform into two single Weyl points when the $C$$_3$ symmetry is broken. Such a scenario can be realized by applying uniaxial lattice strains in the \emph{a-b} plane [see Fig.~\ref{fig6}(d)]. In Fig.~\ref{fig6}(e), we show the transformation of $W$$_1$ from a double-Weyl point into two single-Weyl points under a 2$\%$ compression of uniaxial strain. Considering the slopes of the crossing bands, the single-Weyl points belong to the type-II category. In K$_2$RhF$_6$ compound, the form of Weyl fermions can be mutually shifted by strain engineering, which provides a tunable platform for studying the exotic properties of various Weyl states.

\section{Remarks and conclusion}

We have following remarks before closing. First, in the above presentation, we take K$_2$RhF$_6$ as a representative example. We have calculated the electronic band structure for other members of X$_2$RhF$_6$ (X=K, Rb, Cs) compounds~\cite{63}. We find that the essential features, including the half-metallic band structure and type-III double-Weyl states, are also present. In these compounds, the band gap in the spin-up channel all above 2 eV, which ensure the 100$\%$ spin-polarized conducting electrons. Meanwhile, the double-Weyl points in X$_2$RhF$_6$ compounds all locate close to the Fermi level, and have ``clean" band structures without other rambling bands nearby. These features can be advantages for the experimental detections of the double-Weyl states in these materials.

Second, the double Weyl points proposed here are solely formed by the bands from the spin-down channel; thus, they are fully spin-polarized. Fully spin-polarized fermions have also been proposed in several other half-metal phases, such as the nodal-loop fermions in compounds Li$_3$(FeO$_3$)$_2$~\cite{69}, $\beta$-V$_2$OPO$_4$~\cite{70}, MnN and CrN monolayers ~\cite{71,72}, nodal-chain fermions in half-metallic LiV$_2$O$_4$~\cite{73}, Dirac/Weyl fermions in PtCl$_3$ monolayer~\cite{74}, YN$_2$ monolayer~\cite{75,76}, and some spinel compounds~\cite{77}. Such fully spin-polarized fermions are believed for special spintronics applications~\cite{69,71}. Previously, double-Weyl fermions are mostly proposed in nonmagnetic materials~\cite{50,52,53,54}. Half metal HgCr$_2$Se$_4$ is a typical candidate for fully spin-polarized double-Weyl fermions~\cite{38}. However, the double-Weyl fermions in HgCr$_2$Se$_4$ have different dispersions with X$_2$RhF$_6$ compounds proposed here, where HgCr$_2$Se$_4$ shows type-I dispersion while those in X$_2$RhF$_6$ show type-III. To the best our knowledge, (TaSe$_4$)$_2$I compound is the only candidate for type-III double-Weyl fermions so far~\cite{35}, which provides a nonmagnetic platform to investigate their properties. However, the type-III double-Weyl fermions in X$_2$RhF$_6$ compounds can tell a different story, because these fermions are present in magnetic system and are even fully spin-polarized.

In conclusion, we have demonstrated quasi-one dimensional materials X$_2$RhF$_6$ (X=K, Rb, Cs) as the first material platform to realize fully spin-polarized double-Weyl states with type-III dispersions. By using K$_2$RhF$_6$ compound as an example, we reveal these materials naturally possess half-metallic band structure with large spin gaps in the spin-up channel. In the spin-down channel, when SOC is not included, the band crossing forms a pair of TDNPs in the $\Gamma$-A path near the Fermi level. After SOC is taken into account, the TDNPs splits into two pairs of double-Weyl points, which feature quadratic dispersion along two momentum directions and have the chiral charge of $\pm$2. These double-Weyl points are formed by the crossing bands from the same spin channel, thus they are fully spin-polarized. Even interestingly, one of the crossing bands for the Weyl points has a saddle shape, making the Weyl points belong to type-III. Such fully spin-polarized double-Weyl fermions with type-III dispersion have not been identified in realistic materials before. In addition, we show the double-Weyl fermions in K$_2$RhF$_6$ are characterized with two pieces of Fermi arcs on the surface spectrum. The Weyl points are protected by the $C$$_3$ symmetry. We show the position of Weyl points can be shifted under biaxial strain with $C$$_3$ symmetry preserved. Breaking the $C$$_3$ symmetry by uniaxial strain can transform the double-Weyl points into pairs of single-Weyl points with type-II dispersion. Our results reveal that these X$_2$RhF$_6$ materials are excellent candidates to investigate the novel properties of fully spin-polarized double-Weyl fermions with type-III dispersion.

\begin{acknowledgments}

This work is supported by National Natural Science Foundation of China (Grants No. 11904074), Nature Science Foundation of Hebei Province (No. E2019202222 and E2019202107). One of the authors (X.M. Zhang) acknowledges the financial support from Young Elite Scientists Sponsorship Program by Tianjin.

\end{acknowledgments}


\begin{thebibliography}{}
\bibitem{1} K. S. Novoselov, A. K. Geim, S. V. Morozov, D. Jiang, Y. Zhang, S. V. Dubonos, I. V. Grigorieva, and A. A. Firsov, Science \textbf{306}, 666 (2004).
\bibitem{2} A. Castro Neto, F. Guinea, N. M. Peres, K. S. Novoselov, and A. K. Geim, Rev. Mod. Phys. \textbf{81}, 109 (2009).
\bibitem{3} K. S. Novoselov, A. Mishchenko, A. Carvalho, and A. H. Castro Neto, Science \textbf{353}, aac9439 (2016).
\bibitem{4} M. Hasan and C. L. Kane, Rev. Mod. Phys. \textbf{82}, 3045 (2010).
\bibitem{5} X. L. Qi and S. C. Zhang, Rev. Mod. Phys. \textbf{83}, 1057 (2011).
\bibitem{6} X. G. Wan, A. M. Turner, A. Vishwanath, and S. Savrasov, Phys. Rev. B \textbf{83}, 205101 (2011).
\bibitem{7} S. Murakami, New. J. Phys. \textbf{9}, 356 (2007).
\bibitem{8} A. A. Burkov, Nat. Mater. \textbf{15}, 1145 (2016).
\bibitem{9} A. Bansil, H. Lin, and T. Das, Rev. Mod. Phys. \textbf{88}, 021004 (2016).
\bibitem{10} N. P. Armitage, E. J. Mele, and A. Vishwanath, Rev. Mod. Phys. \textbf{90}, 015001 (2018).
\bibitem{11} X. Huang, L. Zhao, Y. Long, P. Wang, D. Chen, Z. Yang, H. Liang, M. Xue, H. Weng, Z. Fang, X. Dai, and G. Chen, Phys. Rev. X \textbf{5}, 031023 (2015).
\bibitem{12} J. Liu and D. Vanderbilt, Phys. Rev. B \textbf{90}, 155316 (2014).
\bibitem{13} M. Hirayama, R. Okugawa, S. Ishibashi, S. Murakami, and T. Miyake, Phys. Rev. Lett. \textbf{114}, 206401 (2015).
\bibitem{14} D. T. Son and B. Z. Spivak, Phys. Rev. B \textbf{88}, 104412 (2013).
\bibitem{15} C.-X. Liu, P. Ye, and X.-L. Qi, Phys. Rev. B \textbf{87}, 235306 (2013).
\bibitem{16} S. M. Young, S. Zaheer, J. C. Y. Teo, C. L. Kane, E. J.Mele, and A. M. Rappe, Phys. Rev. Lett. \textbf{108}, 140405 (2012).
\bibitem{17} H. Weng, C. Fang, Z. Fang, B. A. Bernevig, and X. Dai, Phys. Rev. X \textbf{5}, 011029 (2015).
\bibitem{18} G. Xu, H. Weng, Z. Wang, X. Dai, and Z. Fang, Phys. Rev. Lett. \textbf{107}, 186806 (2011).
\bibitem{19} C. Fang, M. J. Gilbert, X. Dai, and B. A. Bernevig, Phys. Rev. Lett. \textbf{108}, 266802 (2012).
\bibitem{20} L. Lepori, I. C. Fulga, A. Trombettoni, and M. Burrello, Phys. Rev. A \textbf{94}, 053633 (2016).
\bibitem{21} S.-M. Huang, S.-Y. Xu, I. Belopolski, C.-C. Lee, G. Chang, T.-R. Chang, B. Wang, N. Alidoust, G. Bian, M. Neupane, D. Sanchez, H. Zheng, H.-T. Jeng, A. Bansil, T. Neupert, H. Lin, and M. Z. Hasan, Proc. Natl. Acad. Sci. U.S.A. \textbf{113}, 1180 (2016).
\bibitem{22} S. S. Tsirkin, I. Souza, and D. Vanderbilt, Phys. Rev. B \textbf{96}, 045102 (2017).
\bibitem{23} B. Q. Lv, H. M. Weng, B. B. Fu, X. P. Wang, H. Miao, J. Ma, P. Richard, X. C. Huang, L. X. Zhao, G. F. Chen, Z. Fang, X. Dai, T. Qian, and H. Ding, Phys. Rev. X  \textbf{5}, 031013 (2015).
\bibitem{24} S. M. Huang, S. Y. Xu, I. Belopolski, C. C. Lee, G. Chang, B. Wang, N. Alidoust, G. Bian, M. Neupane, C. Zhang, S. Jia, A. Bansil, H. Lin, and M. Z. Hasan, Nat. Commun. \textbf{6}, 7373 (2015).
\bibitem{25} S. Y. Xu, I. Belopolski, N. Alidoust, M. Neupane, G. Bian, C. Zhang, R. Sankar, G. Chang, Z. Yuan, C.-C. Lee, S.-M. Huang, H. Zheng, J. Ma, D. S. Sanchez, B. Wang, A. Bansil, F. Chou, P. P. Shibayev, H. Lin, S. Jia \emph{et al}., Science \textbf{349}, 613 (2015).
\bibitem{26} T.-R. Chang, S.-Y. Xu, G. Chang, C.-C. Lee, S.-M. Huang, B. G. Wang, H. Bian, D. S. Zheng, I. Sanchez \emph{et al}., Nat. Commun. \textbf{7}, 10639 (2016).
\bibitem{27} A. A. Soluyanov, D. Gresch, Z. Wang, Q. Wu, M. Troyer, X. Dai, and B. A. Bernevig, Nature \textbf{527}, 495 (2015).
\bibitem{28} T.-R. Chang, S.-Y. Xu, D. S. Sanchez, W.-F. Tsai, S. M. Huang, G. Chang, C. H. Hsu, G. Bian, I. Belopolski, Z.-M. Yu \emph{et al}., Phys. Rev. Lett. \textbf{119}, 026404 (2017).
\bibitem{29} A. A. Soluyanov, D. Gresch, Z. Wang, Q. Wu, M. Troyer, X. Dai, and B. A. Bernevig, Nature \textbf{527}, 495 (2015).
\bibitem{30} H. Isobe and N. Nagaosa, Phys. Rev. Lett. \textbf{116}, 116803 (2016).
\bibitem{31} G. Chang, S.-Y. Xu, D. S. Sanchez, S.-M. Huang, C.-C. Lee, T.-R. Chang, G. Bian, H. Zheng, I. Belopolski, N. Alidoust, H.-T. Jeng, A. Bansil, H. Lin, and M. Zahid Hasan, Sci. Adv. \textbf{2}, e1600295 (2016).
\bibitem{32} Z.-M. Yu, Y. Yao, and S. A. Yang, Phys. Rev. Lett. \textbf{117}, 077202 (2016).
\bibitem{33} T. E. O'Brien, M. Diez, and C. W. J. Beenakker, Phys. Rev. Lett. \textbf{116}, 236401 (2016).
\bibitem{34} Z.-M. Yu, W. K. Wu, X.-L. Sheng, Y. X. Zhao, and S. Y. A. Yang, Phys. Rev. B \textbf{99}, 121106(R) (2019).
\bibitem{35} X.-P. Li, K. Deng, B. T. Fu, Y. K. Li, D. S. Ma, J. F. Han, J. H. Zhou, S. Y. Zhou, and Y. G. Yao, arXiv:1909.12178v1.
\bibitem{36} H. M. Weng, C. Fang, Z. Fang, B. Andrei Bernevig, and X. Dai, Phys. Rev. X \textbf{5}, 011029 (2015).
\bibitem{37} B. Q. Lv, H. M. Weng, B. B. Fu, X. P. Wang, H. Miao, J. Ma, P. Richard, X. C. Huang, L. X. Zhao, G. F. Chen, Z. Fang, X. Dai, T. Qian, and H. Ding, Phys. Rev. X  \textbf{5}, 031013 (2015).
\bibitem{38} C. Fang, M. J. Gilbert, X. Dai, and B. Andrei Bernevig, Phys. Rev. Lett. \textbf{108}, 266802 (2012).
\bibitem{39} G. Q. Chang, B. Singh, S.-Y. Xu, G. Bian, S.-M. Huang, C.-H. Hsu, I. Belopolski, N. Alidoust, D. S. Sanchez, H. Zheng, H. Lu, X. Zhang, Y. Bian, T.-R. Chang, H.-T. Jeng, A. Bansil, H. Hsu, S. Jia, T. Neupert, H. Lin, and M. Zahid Hasan, Phys. Rev. B \textbf{97}, 041104(R) (2018).
\bibitem{40} D. S. Sanchez, G. Q. Chang, I. Belopolski, H. Lu, J.-X. Yin, N. Alidoust, X. T. Xu, T. A. Cochran, X. Zhang, Y. Bian, S. T. S. Zhang, Y.-Y. Liu, J. Ma, G. Bian, H. Lin, S.-Y. Xu, S. Jia, and M. Zahid Hasan, Nat. Commun. \textbf{11}, 3356 (2020).
\bibitem{41} C. Y. Huang, H. Aramberri, H. Lin, and N. Kioussis, arXiv:2005.08914.
\bibitem{42} Z. J. Wang, M. G. Vergniory, S. Kushwaha, M. Hirschberger, E. V. Chulkov, A. Ernst, N. P. Ong, R. J. Cava, and B. Andrei Bernevig, Phys. Rev. Lett. \textbf{117}, 236401 (2016).
\bibitem{43} E. Liu, Y. Sun, N. Kumar, L. Muechler, A. Sun, Li. Jiao, S.-Y. Yang, D. Liu, A. Liang, Q. Xu, J. Kroder, V. S\"{u}{\ss}, H. Borrmann, C. Shekhar, Z. S. Wang, C. Xi, W. Wang, W. Schnelle, S. Wirth, Y. Chen, S. T. B. Goennenwein, and C. Felser, Nat. Phys. \textbf{14}, 1125 (2018).
\bibitem{44} L. Muechler, E. Liu, Q. Xu, C. Felser, and Y. Sun, arXiv:1712.08115v2.
\bibitem{45} S. Han, C. Lee, E.-G. Moon, and H. Min, Phys. Rev. Lett. \textbf{122}, 187601 (2019).
\bibitem{46} J.-R. Wang, G.-Z. Liu, and C.-J. Zhang, Phys. Rev. B \textbf{99}, 195119 (2019).
\bibitem{47} S.-X. Zhang, S.-K. Jian, and H. Yao, arXiv:1809.10686.
\bibitem{48} Q. Chen and G. A. Fiete, Phys. Rev. B \textbf{93}, 155125 (2016).
\bibitem{49} H.-F. Zhu, X.-Q. Yang, J. Xu, and S. Cao, Eur. Phys. J. B \textbf{93}, 4 (2020).
\bibitem{50} S.-M. Huanga, S.-Y. Xuc, I. Belopolskic, C.-C. Leea, G. Changa, T.-R. Changc, B. K. Wanga, N. Alidoustc, G. Bianc, M. Neupanec, D. Sanchezc, H. Zhengc, H.-T. Jengd, A. Bansilf, T. Neuperth, H. Lina, and M. Zahid Hasan, PNAS \textbf{113}, 1180 (2016).
\bibitem{51} H. B. Wu, D.-S. Ma, B. T. Fu, W. Guo, and Y. G. Yao, J. Phys. Chem. Lett. \textbf{10}, 2508 (2019).
\bibitem{52} T. T. Zhang, Z. D. Song, A. Alexandradinata, H. M. Weng, C. Fang, L. Lu, and Z. Fang, Phys. Rev. Lett. \textbf{120}, 016401 (2018).
\bibitem{53} D. Takane, Z. Wang, S. Souma, K. Nakayama, T. Nakamura, H. Oinuma, Y. Nakata, H. Iwasawa, C. Cacho, T. Kim, K. Horiba, H. Kumigashira, T. Takahashi, Y. Ando, and T. Sato, Phys. Rev. Lett. \textbf{122}, 076402 (2019).
\bibitem{54} L. Y. Zhu, S.-S. Wang, S. Guan, Y. Liu, T. T. Zhang, G. B. Chen, and S. Y. A. Yang, Nano Lett. \textbf{16}, 6548 (2016).
\bibitem{55} G. Kresse and D. Joubert, Phys. Rev. B: Condens. Matter Mater. Phys. \textbf{59}, 1758 (1999).
\bibitem{56} P. E. Blochl, Phys. Rev. B: Condens. Matter Mater. Phys. \textbf{50}, 17953 (1994).
\bibitem{57} J. P. Perdew, K. Burke, and M. Ernzerhof, Phys. Rev. Lett. \textbf{77}, 3865 (1996).
\bibitem{58} P. Bl\"{o}chl, Phys. Rev. B \textbf{50}, 17953 (1994).
\bibitem{59} N. Marzari and D. Vanderbilt, Phys. Rev. B: Condens. Matter Mater. Phys. \textbf{56}, 12847 (1997).
\bibitem{60} A. A. Mostofi, J. R. Yates, Y.-S. Lee, I. Souza, D. Vanderbilt, and N. Marzari, Comput. Phys. Commun. \textbf{178}, 685 (2008).
\bibitem{61} Q. S. Wu, S. N. Zhang, H.-F. Song, M. Troyer, and A. A. Soluyanov, Comput. Phys. Commun. \textbf{224}, 405 (2018).
\bibitem{62} E. Weise and W. Klemm, Zeitschrift fuer Anorganische und Allgemeine Chemie \textbf{272}, 211 (1953).
\bibitem{63} See supplementary Information for the considered magnetic configurations and density of states for K$_2$RhF$_6$ compound, and electronic band structures for Rb$_2$RhF$_6$ and Cs$_2$RhF$_6$ compounds.
\bibitem{64} Z. Zhu, G. W. Winkler, Q. S. Wu, J. Li, and A. A. Soluyanov, Phys. Rev. X \textbf{6}, 031003 (2016).
\bibitem{65} H. Weng, C. Fang, Z. Fang, and X. Dai, Phys. Rev. B, \textbf{93}, 241202 (2016).
\bibitem{66} H. Weng, C. Fang, Z. Fang, and X. Dai, Phys. Rev. B \textbf{94}, 165201 (2016).
\bibitem{67} X. M. Zhang, Z.-M. Yu, X.-L. Sheng, H. Y. Yang, and S. Y. A. Yang, Phys. Rev. B \textbf{95}, 235116 (2017).
\bibitem{68} L. Jin, X. M. Zhang, X. F. Dai, H. Y. Liu, G. F. Chen, and G. D. Liu, J. Mater. Chem. C \textbf{7}, 1316 (2019).
\bibitem{69} C. Chen, Z.-M. Yu, S. Li, Z. Chen, X.-L. Sheng, and S. Y. A. Yang, arXiv:1811.05254v1.
\bibitem{70} Y. J. Jin, R. Wang, Z. J. Chen, J. Z. Zhao, Y. J. Zhao, and H. Xu, Phys. Rev. B \textbf{96}, 201102(R) (2017).
\bibitem{71} S.-S. Wang, Z.-M. Yu, Y. Liu, Y. Jiao, S. Guan, X.-L. Sheng, and S. Y. A. Yang, Phys. Rev. Mater. \textbf{3}, 084201 (2019).
\bibitem{72} T. L. He, X. M. Zhang, Z.-M. Yu, Y. Liu, X. F. Dai, G. D. Liu, and Y. G. Yao,  arXiv:2007.12839.
\bibitem{73} H. P. Zhang, X. M. Zhang, Y. Liu, X. F. Dai, G. Chen, and G. D. Liu, arXiv:2007.12842.
\bibitem{74} J.-Y. You, C. Chen, Z. Zhang, X.-L. Sheng, S. A. Yang, and G. Su, Phys. Rev. B \textbf{100}, 064408 (2019).
\bibitem{75} Z. F. Liu, J. Y. Liu, and J. J. Zhao, Nano Res. \textbf{10}, 1972 (2017).
\bibitem{76} X. Kong, L. Li, O. Leenaerts, W. Wang, X.-J. Liua, and F. M. Peeters, Nanoscale \textbf{10}, 8153 (2018).
\bibitem{77} W. Jiang, H. Huang, F. Liu, J.-P. Wang, and T. Low, Phys. Rev. B \textbf{101}, 121113(R) (2020).



\end{thebibliography}
\end{document}